\documentclass[pra,twocolumn,superscriptaddress]{revtex4}
\usepackage{amsbsy}
\usepackage{latexsym,epsfig,graphicx}
\usepackage{dcolumn}
\usepackage{graphicx}
\usepackage{subfigure}
\usepackage{comment}
\usepackage{color}
\usepackage{bm}
\usepackage{mathrsfs}
\usepackage{amsfonts}
\usepackage{amsmath}
\usepackage{color}
\usepackage{amssymb}
\usepackage{xspace}
\usepackage{epstopdf}
\usepackage{tabularx}
\usepackage{longtable}
\usepackage[colorlinks=true, letterpaper=true, pdfstartview=FitV, linkcolor=blue, citecolor=blue, urlcolor=blue]{hyperref}
\usepackage[normalem]{ulem}

\pdfoutput=1
\input{tcilatex}
\begin{document}

\title{Symmetry-protected localized states at defects in non-Hermitian
systems}
\author{Ya-Jie Wu}
\thanks{wuyajie@xatu.edu.cn}
\affiliation{School of Science, Xi'an Technological University, Xi'an 710032, China}
\affiliation{Department of Physics, The University of Texas at Dallas, Richardson, Texas
75080-3021, USA}
\author{Junpeng Hou}
\thanks{Junpeng.Hou@utdallas.edu}
\affiliation{Department of Physics, The University of Texas at Dallas, Richardson, Texas
75080-3021, USA}

\begin{abstract}
Understanding how local potentials affect system eigenmodes is crucial for
experimental studies of nontrivial bulk topology. Recent studies have
discovered many exotic and highly non-trivial topological states in
non-Hermitian systems. As such, it would be interesting to see how
non-Hermitian systems respond to local perturbations. In this work, we
consider chiral and particle-hole -symmetric non-Hermitian systems on a
bipartite lattice, including SSH model and photonic graphene, and find that
a disordered local potential could induce bound states evolving from the
bulk. When the local potential on a single site becomes infinite, which
renders a lattice vacancy, chiral-symmetry-protected zero-energy mode and
particle-hole symmetry-protected bound states with purely imaginary
eigenvalues emerge near the vacancy. These modes are robust against any
symmetry-preserved perturbations. Our work generalizes the
symmetry-protected localized states to non-Hermitian systems.
\end{abstract}

\maketitle

\section{Introduction and Motivation}

Non-Hermitian Hamiltonian captures essentials of open systems governed by
non-Hermitian operators \cite{Carmichael1993,Rotter2009,Choi2010,Diehl2011,
Lee2014a,Lee2014b,Malzard2015,Zhen2015,Cao2015,Jose2016}, for instance,
optical and mechanical structures with gain and loss \cite{Makris2008,
Chong2011,Regensburger2012,Hodaei2014,Peng2014a,Feng2014,Jing2014,
Peng2014b,liu2014,liu2016,Kawabata2017,Ashida2017,Weimann2017,Ganainy2018}.
Intriguingly, although non-Hermitian operators usually have complex
eigenvalues, the energy spectrums of a non-Hermitian Hamiltonian with
parity-time ($\mathcal{PT}$) symmetry could be real-valued in $\mathcal{PT}$%
-symmetric regimes. Such an reality could also be broken by tuning, for
example, gain/loss strength, and in the resulted $\mathcal{PT}$-broken
regime, the $\mathcal{PT}$ symmetry is said to be broken spontaneously \cite%
{Bender1998,Bender2007}. $\mathcal{PT}$ symmetry breaking has already been
observed in optical waveguides \cite{Ruter2010}. Similar physics exist in $%
\mathcal{CP}$ symmetry, where $\mathcal{C}$ denotes particle-hole symmetry,
due to the anti-linearity of $\mathcal{C}$ and $\mathcal{T}$. For a $%
\mathcal{CP}$-symmetric Hamiltonian $H$, $\mathcal{CP}$ and $\mathcal{PT}$
symmetries are equivalent under the transformation $H\rightarrow iH$ \cite%
{Kawabata2019,Ryo2019,Yamamoto2019}. Consequently, the eigenenergies of a $%
\mathcal{CP}$-symmetric system is imaginary when $\mathcal{CP}$ symmetry is
preserved in the spectrum. Otherwise, it could be real in the $\mathcal{CP}$%
-broken regimes.

On the other hand, topological states have attracted intensive attentions in
various Hermitian systems \cite{Hansan2010, Qix2011}. Recently, the concept
of topological phases have been extended to non-Hermitian systems. $\mathcal{%
C}$ and $\mathcal{T}$ symmetries are unified by non-Hermiticity, which
allows topological phases in high dimensions. The interplay between topology
and non-Hermiticity leads to rich topological features with no Hermitian
counterpart \cite{Rudner2009,
Liang2013,Zhu2014,Leykam2017,Gong2017,Gonz2017,Shen2018,
Lieu2018,Yin2018,Li2018,Yao2018a,Yao2018b,Gong2018a,Kawabata2018,Harari2018}%
. In particular, the conventional bulk-boundary correspondence breaks down
in non-Hermitian systems and new topological invariants like non-Bloch
topological invariant and vorticity must be introduced to understand the
underlying topological properties.

The nontrivial bulk topology in Hermitian systems can be detected by
defects, such as edges, $\pi $-flux, dislocations and vortices \cite%
{Weeks2007,Tewari2007,Rosenberg2009,Roy2010,Santos2011,Juricic2012}. When it
comes to non-Hermitian systems, stable edge states could also exist at the
interface between topological and trivial phases \cite%
{Yuce2016,Yuce2018a,Lang2018,Yuce2018b,Oztas2018,Jan2019,Okugawa2019,Tsuneya2019,Hengyun2019}%
. These topological states, originated from bulk topologies, are immune to
local symmetry-preserved perturbations. It is well known that a local
potential could induce localized modes in topological phases of Hermitian
systems \cite{Shan2011,Balatsky2006,Lu2011}, while such a problem has far
less been investigated in non-Hermitian systems. In addition, recent studies
of topological states in open systems have found many novel and unique
topological phases in non-Hermitian systems. In this sense, it is worth
investigating how a local potential affect system eigenmodes in
non-Hermitian systems. In general, for a bipartite lattice with Hamiltonian $%
H$ obeying the symmetry $\mathcal{O}H\mathcal{O}^{-1}=-H$, the quantum
states are paired with opposite real parts of eigenvalues. Then, once a
single lattice site is removed by an infinite local potential, an unpaired
mode with zero or purely imaginary energy appears.

In this work, we generalize the idea to non-Hermitian systems and show the
robustness of the induced bound states. Specifically, we focus on both
1-dimensional (1D) and 2D systems with two sublattice degrees of freedoms,
respecting either chiral ($\mathcal{O}=\mathcal{S}$) or particle-hole ($%
\mathcal{O}=\mathcal{C}$) symmetry, which are responsible for versatile symmetry-protected topological phases in low dimensions. We show that, in the cases of
non-Hermitian systems, the lattice vacancy can induce symmetry-protected
localized modes in both topological and trivial phases.

The remaining of this paper is organized as following. In Sec. \ref{S2}, we
discuss $\mathcal{S}$ and $\mathcal{C}$ symmetries on a bipartite lattice,
and derive the eigenvalue-correspondence relation. We start with a 1D system
in Sec. \ref{S3}, namely, the non-Hermitian Su-Schrieffer-Heeger (SSH) model
with either $\mathcal{S}$ or $\mathcal{C}$ symmetry, and study the effects
of a lattice vacancy. In Sec. \ref{S4}, we extend to a 2D photonic graphene.
We apply both symmetry analysis and numerical calculations to investigate
how lattice vacancies change the system eigenmodes. Finally, conclusions and
discussions are presented in Sec. \ref{S5}.

\section{Chiral and particle-hole symmetries in non-Hermitian systems}

\label{S2} In this section, we study the general theory of
symmetry-protected modes induced by vacancies. For simplicity, we consider
non-Hermitian effective models on a bipartite lattice (sublattice \textrm{A}
and \textrm{B}). In momentum space, the generic effective Hamiltonian is $%
\hat{H}=\sum_{k}\Psi _{k}^{\dagger}H\left( k\right) \Psi _{k}$ with $\hat{%
\Psi}_{k}^{\dagger }=\left( \hat{a}_{k}^{\dagger },\hat{b}_{k}^{\dagger
}\right) $ and%
\begin{equation}
H\left( k\right) =\bm{\mathit{h}}_{0}\cdot \mathbf{\sigma }+i\bm{\mathit{h}}%
_{1}\cdot \mathbf{\sigma ,}
\end{equation}%
where $\sigma _{0}$ and $\mathbf{\sigma =}\left( \sigma _{x},\sigma
_{y},\sigma _{z}\right) $ are identity matrix and Pauli matrices that act on
sublattice space respectively and $\bm{\mathit{h}}_{i}=\left(
h_{i,x},h_{i,y},h_{i,z}\right), i=0,1$ are real.

Firstly, we consider $\mathcal{S}$ symmetry described by $\mathcal{S}H\left(
k\right)\mathcal{S}^{-1}=-H\left( k\right) $, where $S$ is a unitary
operator. When $S=\sigma_{z}$ is chosen in this basis, we obtain
\begin{equation}
H\left( k\right) =h_{0,x}\sigma _{x}+h_{0,y}\sigma _{y}+ih_{1,x}\sigma
_{x}+ih_{1,y}\sigma _{y}.
\end{equation}
If $\psi _{k}$ is an eigenstate for Hamiltonian $H\left( k\right) $ with
eigenvalue $E_{k}$, $\mathcal{S}\psi _{k}$ is an eigenstate for Hamiltonian $%
H\left(k\right) $ with eigenvalue $-E_{k}$. Thus, for above non-Hermitian
system on a bipartite lattice, there exists following energy-eigenvalue
correspondence: $E_{k}\Leftrightarrow -E_{k}$. This symmetry dictates energy
eigenvalues must be paired.

\begin{figure}[tbp]
\centering\includegraphics[width=0.48\textwidth]{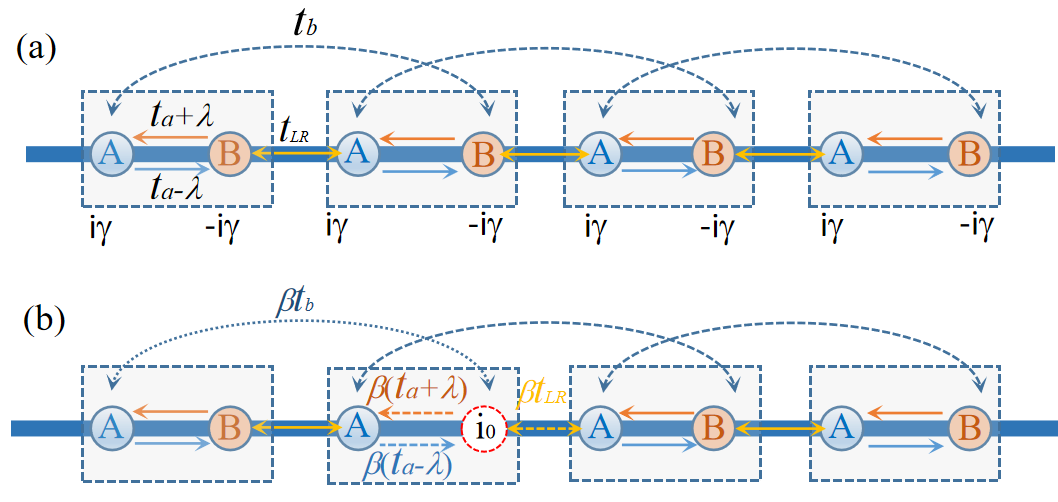}
\caption{Illustration of non-Hermitian SSH model. The dotted rectangle
denotes the unit cell. (a) $t_{a}\pm \protect\lambda $, $t_{LR}$ and $t_{b}$
are tunneling strength and $\pm i\protect\gamma $ denote balanced gain and
loss. (b) The red dashed circle at $i_{0}$ denotes the site under a local
potential $V_{d}=V_{0}\left( 1-\protect\beta \right) /\protect\beta $. The
hopping amplitude related to this site is proportional to $\protect\beta $.}
\label{Fig1}
\end{figure}

Secondly, let us consider $\mathcal{C}$ symmetry described by $\mathcal{C}%
H\left( k\right)\mathcal{C}^{-1}=-H\left( -k\right) $ and $\mathcal{C}$
symmetry being antiunitary \cite{Note1}. When $\mathcal{C}=\sigma_{z}K$ is
chosen in this basis, we have
\begin{equation}
H\left( k\right) =h_{0,x}\sigma _{x}+h_{0,y}\sigma _{y}+ih_{1,z}\sigma _{z}
\end{equation}%
with constraints $h_{0,x}\left( k\right) =h_{0,x}\left( -k\right) $, $%
h_{0,y}\left(k\right) =-h_{0,y}\left( -k\right) $ and $h_{1,z}\left(
k\right)=h_{1,z}\left( -k\right) $. Provided that $\psi _{k}$ is an
eigenstate with eigenvalue $E_{k}$ for Hamiltonian $H\left( k\right) $, $%
\mathcal{C}\psi _{k}$ is an eigenstate of Hamiltonian $H\left( -k\right) $
with eigenvalue $-E_{k}^{\ast }$. Therefore, the energy spectra has
correspondence $E_{k}\Leftrightarrow -E_{-k}^{\ast }$ under periodic
boundary condition. This symmetry classifies energy eigenvalues in
complex-conjugate pairs, except when they are purely imaginary.

Consider a $\mathcal{S}$-symmetric non-Hermitian system with $N_{\mathrm{u}}$
unitcells. If a lattice site is removed (corresponds to a lattice vacancy
defect), the translation symmetry is broken, but $\mathcal{S}$ symmetry of
the Hamiltonian is still preserved through the transformations $\hat{a}%
_{i}\Rightarrow \hat{a}_{i}$, $\hat{b}_{i}\Rightarrow -\hat{b}_{i}$ and $%
\hat{H}\Rightarrow -\hat{H}$, where $\hat{a}_{i}\ $($\hat{b}_{i}$) denotes
annihilation operators on lattice site $i$ of \textrm{A} (\textrm{B})
sublattice. Now, only $2N_{\mathrm{u}}-1$ quantum states are available. It
leads to the energy-eigenvalue correspondence $E_{1,...,N_{\mathrm{u}%
}-1}\Leftrightarrow -E_{N_{\mathrm{u}}+1,...,2N_{\mathrm{u}}-1}$, i.e., only
$2N_{\mathrm{u}}-2$ states are paired. To guarantee $\mathcal{S}$ symmetry,
the single left unpaired state must satisfy $E_{N_{\mathrm{u}%
}}\Leftrightarrow -E_{N_{\mathrm{u}}}$. It means the remained single state
must have \textit{zero eigenenergy}. While this argument is the same for
Hermitian and non-Hermitian systems, the physics is richer with
non-Hermiticity as we will see later.

Next, we consider a $\mathcal{C}$-symmetric non-Hermitian system with $N_{%
\mathrm{u}}$ unit cells. When a single lattice site is removed, a lattice
vacancy arises and $2N_{\mathrm{u}}-1$ quantum states remain. At this time, $%
\mathcal{C}$ symmetry of the system is also respected through the
particle-hole transformation $\hat{a}_{i}\Rightarrow \hat{a}_{i}^{\dagger }$%
, $\hat{b}_{i}\Rightarrow -\hat{b}_{i}^{\dagger }$, $\hat{H}\Rightarrow -%
\hat{H}$. It leads to the energy-eigenvalue correspondence $E_{1,...,N_{%
\mathrm{u}}-1}\Leftrightarrow -E_{N_{\mathrm{u}}+1,...,2N_{\mathrm{u}%
}-1}^{\ast }$, i.e., $2N_{\mathrm{u}}-2$ states are conjugate paired. To
guarantee $C$ symmetry, the single left unpaired state must satisfy $E_{N_{%
\mathrm{u}}}\Leftrightarrow -E_{N_{\mathrm{u}}}^{\ast }$. It means this
single unpaired state has a either \textit{zero} or \textit{purely imaginary}
energy. Obviously, the latter is only feasible in non-Hermitian systems.

In the following, we shall provide two concrete examples to elucidate both $%
\mathcal{S}$- and $\mathcal{C}$-symmetry protected modes induced by lattice
vacancy.

\section{Su-Schrieffer-Heeger model}

\label{S3} In this section, we consider the non-Hermitian SSH model shown in
Fig. \ref{Fig1}(a), which is relevant to current experiments. The generic
Bloch Hamiltonian is
\begin{equation}
H_{S,0}\left( k\right) =h_{0,x}\sigma _{x}+\left( h_{0,y}+i\lambda \right)
\sigma _{y}+i\gamma \sigma _{z}.
\end{equation}%
where $h_{0,x}=t_{a}+\left( t_{LR}+t_{b}\right) \cos k,$ $h_{0,y}=\left(
t_{LR}-t_{b}\right) \sin k$. Note that $\ i\lambda \sigma _{y}$ and $i\gamma
\sigma _{z}$ are non-Hermitian parameters, which stem from unequal hopping
strength within a unit cell and balanced gain/loss, respectively. Hereafter,
we will discuss chiral and particle-hole symmetry protected modes induced by
the lattice vacancy.

\subsection{Chiral Symmetry Protected Mode}

\begin{figure}[tbp]
\centering\includegraphics[width=0.46\textwidth]{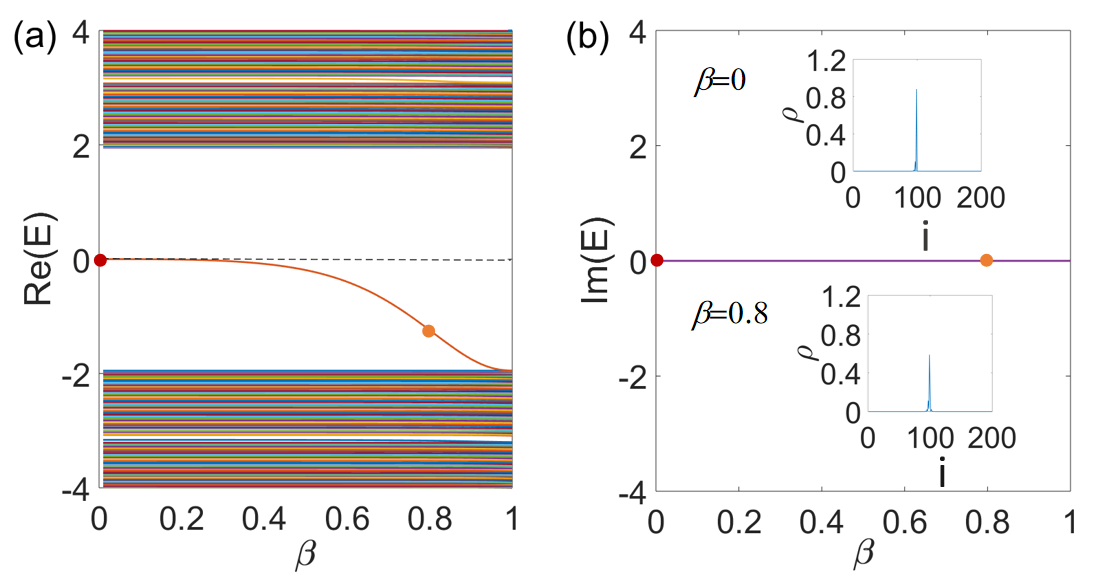}
\caption{Spectrum of SSH model with respect to varying disordered strength $%
\protect\beta $. The real and imaginary parts of eigenvalues are shown in
(a) and (b). The top (bottom) inset shows the particle-density distribution
of localized mode with $\protect\beta =0$ ($\protect\beta =0.8$) in real space. Parameters
are chosen as $t_{a}=1.5$, $\protect\lambda =0.1$, $t_{LR}=1.0$, $t_{b}=0.1$%
, $\protect\gamma =0,V_{0}=10.0$, $N_{u}=100.$}
\label{Fig2}
\end{figure}

When $\gamma =0$, the model has a chiral symmetry $\sigma _{z}H_{S,0}\left(
k\right) \sigma _{z}^{-1}=-H_{S,0}\left( k\right) $. It has been studied in
Ref. \cite{Yao2018a}, where the issue of breakdown of conventional
bulk-boundary correspondence has been settled and non-Bloch bulk-boundary
correspondence was introduced. Chiral symmetry ensures that the eigenvalues
appear in $\left( E_{k},-E_{k}\right) $ pairs. If there exists a vacancy
(see Fig. \ref{Fig1}(b)), the translation symmetry is broken. However, the
chiral symmetry is still respected by the Hamiltonian. Because the SSH model
is based on a bipartite lattice, there exists an unpaired state. Due to the
eigenvalue-correspondence relation discussed in previous section, the
leftover state must have exactly zero energy.
\begin{figure}[tbp]
\centering\includegraphics[width=0.46\textwidth]{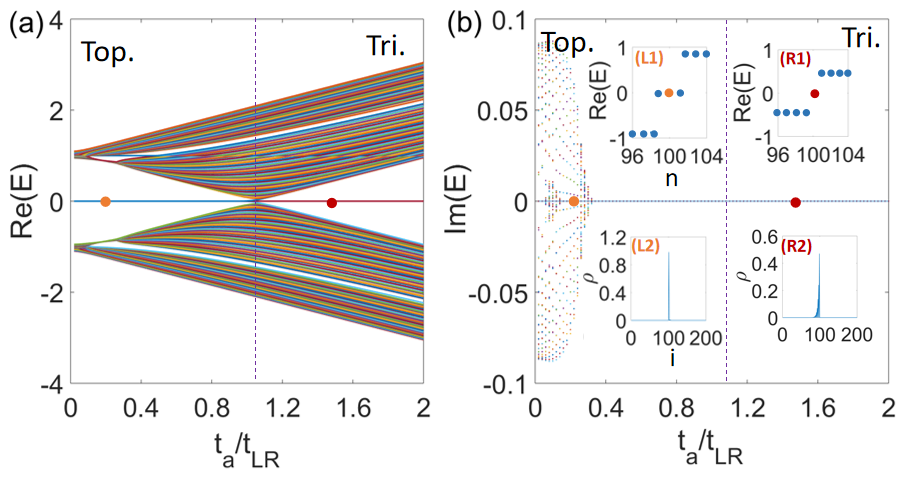}
\caption{The real (a) and imaginary (b) spectrums of SSH model with chiral
symmetry. The orange and red dots indicate two zero-energy states localized
near the vacancy in two topologically distinct phases. The insets (L1) and
(L2) in (b) show the real part of energies $\func{Re}(\mathrm{E})$ versus
the indices of states $\mathrm{n}$ and the particle density distribution $%
\protect\rho $ versus lattice site indices $\mathrm{i}$ of vacancy-induced
zero-energy mode in `Top' phase ($\protect\beta =0.2$, indicated by the
orange dot in (a) and (b)), respectively. The insets (R1) and (R2) present $%
\func{Re}(\mathrm{E})$ and $\protect\rho $ of zero-energy mode in `Tri'
phase ($\protect\beta =1.5$, indicated by the red dot in (a) and (b)),
respectively. Parameters are chosen as $\protect\lambda =0.1$, $t_{LR}=1.0$,
$t_{b}=0.1$, $\protect\gamma =0,V_{0}=10.0$, $N_{u}=100.$ }
\label{Fig3}
\end{figure}

Next, we numerically study the effects of lattice vacancy on the quantum
states within the system. The vacancy can be seen as a \textquotedblleft
hole\textquotedblright\ in the system by removing a lattice site. To
simulate the vacancy, we gradually vary the local potential on a given site $%
i_{0}$ labeled in Fig. \ref{Fig1}(b). The overall Hamiltonian is then $\hat{H%
}_{S,0}=\hat{H}_{S,0}\left( i\neq i_{0}\right) +\hat{H}_{V}$, where $\hat{H}%
_{S,0}\left( i\neq i_{0}\right) $ doesn't contain terms related to the site $%
i_{0}$, and $\hat{H}_{V}$ is
\begin{equation}
\hat{H}_{V}=\beta \sum_{i_{0},j}\left( t_{i_{0},j}\hat{c}_{i_{0}}^{\dagger }%
\hat{c}_{j}+t_{j,i_{0}}\hat{c}_{j}^{\dagger }\hat{c}_{i_{0}}\right)
+\sum_{i_{0}}V_{d}\hat{c}_{i_{0}}^{\dagger }\hat{c}_{i_{0}}.  \label{Eqv}
\end{equation}%
Here, $t_{i_{0},j}$ ($t_{j,i_{0}}$) denotes the bare hopping amplitude
(without local disordered potential) between sites $j$ and $i_{0}$ and the
local potential reads $V_{d}=V_{0}\left( 1-\beta \right) /\beta $. \ When $%
\beta =1$, the local potential $V_{d}=0$. The Hamiltonian $\hat{H}_{V}$
reduces to $\hat{H}_{0}$ and exhibits translation invariance. As $\beta $
decreases, $V_{d}$ gradually increases. When $\beta \rightarrow 0$, the
local potential $V_{d}\rightarrow \infty $ and the effective hopping
amplitude related to site $i_{0}$ approaches zero. This corresponds to a
lattice vacancy at site $i_{0}$. The numerical results are shown in Fig. \ref%
{Fig2}. We see all eigenvalues are real. As $\beta $ decreases, the wave
function evolves from an extended state to an in-gap state. For $0<\beta <1$%
, chiral symmetry is observed to be broken in the spectrums by a bound
state. Such a localized state resides in the energy gap, which is labeled by
the solid tangerine curve in Fig.~\ref{Fig2}(a). When $\beta $ approaches $0$%
, an exact zero-energy state exists, and the energy spectrum becomes
symmetric. The insets of Fig. \ref{Fig2}(b) showcase the particle-density distribution
of localized modes in real space.

\begin{figure}[tbp]
\centering\includegraphics[width=0.46\textwidth]{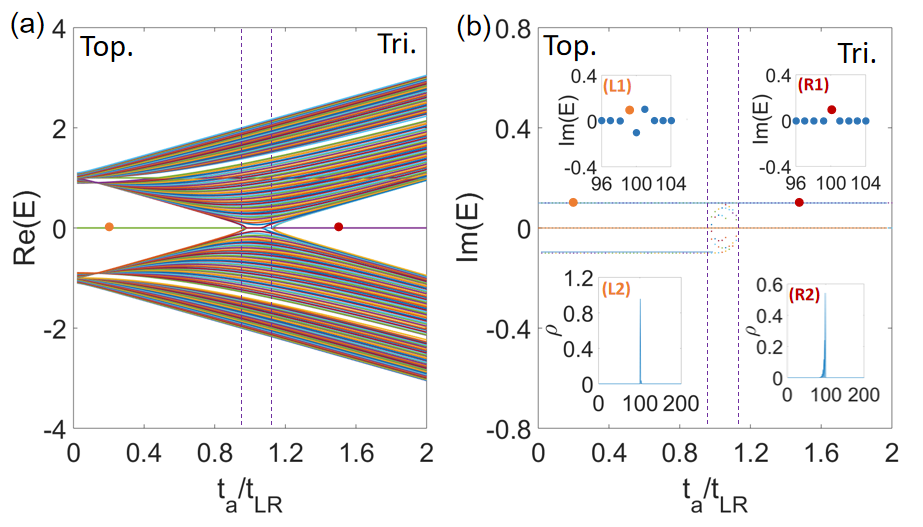}
\caption{Similar as Fig.~\protect\ref{Fig3} but plotted with different
non-Hermitian parameters $\protect\lambda =0$ and $\protect\gamma =0.1$.}
\label{Fig4}
\end{figure}

It is known that there is a topological phase transition by tuning $%
t_{a}/t_{LR}$ \cite{Yao2018a} but the chiral symmetry is always respected.
To see how the vacancy-induced zero-energy modes respond to topological
phase transitions, we choose an open-boundary chain ($2N_{u}$ lattice sites
with a single vacancy) and calculate the eigenvalues at different $%
t_{a}/t_{LR}$. The numerical results are shown in Fig. \ref{Fig3}. There are
two distinct phases, i.e., the topological phase (Top.) and the trivial
phase (Tri.). In the topological phase, besides the two edge states, there
is another zero-energy state localized near the vacancy, as shown in insets
(L1) and (L2) of Fig.~\ref{Fig3}(b). In the trivial phase, the edge states
disappear but the state localized near vacancy survives, as shown in insets
(R1) and (R2) of Fig.~\ref{Fig3}(b). The wave function could be spatially
extended as $t_{b}$ increases, but its energy always remains zero. In
summary, such a chiral-symmetric zero-energy bound state is robust to
topological phase transition.

\subsection{Particle-Hole Symmetry Protected Mode}

When $\lambda =0$, the model has particle symmetry. It ensures the
eigenvalues appear in $\left( E_{k},-E_{k}^{\ast }\right) $ pairs. In
addition, this model also has $\mathcal{PT}$ symmetry $\sigma
_{x}H_{S,0}^{\ast}\left( k\right) \sigma _{x}=H_{S,0}\left( -k\right)$,
consequently, it may possess real spectrum. However, $\mathcal{PT}$ symmetry
could be spontaneously broken in the interval $t_{a}-\gamma
<t_{LR}<t_{a}+\gamma$, leaving complex energies in the spectrum \cite%
{Yuce2018b}. We apply the same methods to simulate the vacancy and study its
effects on the system. The Hamiltonian is $\hat{H}=\hat{H}_{S,0}\left( i\neq
i_{0}\right) +\hat{H}_{V}$, where $\hat{H}_{V}$ is same as Eq. (\ref{Eqv})
except that $V_{d}=i\epsilon _{i_{0}}\gamma /\beta $ with $\epsilon
_{i_{0}\in A}=+1$ and $\epsilon _{i_{0}\in B}=-1$. Obviously, if $\beta =1$,
the system reduces to the Hamiltonian $\hat{H}_{S,0}$ and exhibits
translation symmetry. As $\beta $ decreases, the amplitude for $V_{d}$
increases, but the hopping amplitude related to the site $i_{0}$ decreases.
As $\beta \rightarrow 0$, the effective hopping amplitude from or to the
site $i_{0}$ approaches zero, and $\left\vert V_{d}\right\vert $ becomes
infinite. When $\beta =0$, a lattice vacancy appears at site $i_{0}$.
\begin{figure}[tbp]
\centering\includegraphics[width=0.40\textwidth]{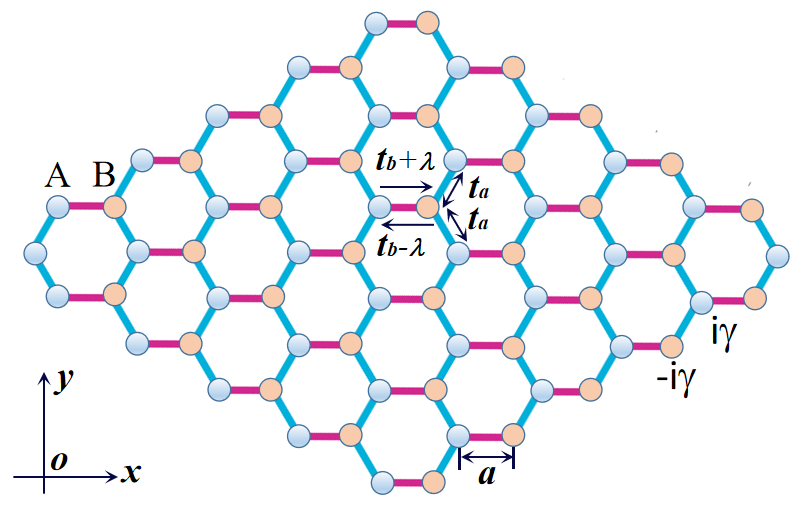}
\caption{Illustration of a honeycomb lattice. Parameters $t_{a}$ and $%
t_{b}\pm \protect\lambda $ are tunneling strength and $\pm i\protect\gamma $
denote balanced gain and loss. The lattice spacing is set as $a=1$.}
\label{Fig5}
\end{figure}

Let us numerically study the system with a single vacancy ($\beta =0$), of
which the translation symmetry is broken, but the particle-hole symmetry is
still respected. Because the SSH model is based on a bipartite lattice,
there exists an unpaired state. Due to the \textquotedblleft spectrum
symmetry\textquotedblright\ ($E\leftrightarrow -E^{\ast }$), the unpaired
state must have exactly \textit{zero} or \textit{purely imaginary } energy.
We calculate eigenenergies for a chain with a single vacancy under open
boundary conditions. The numerical results are shown in Fig. \ref{Fig4} and
there are two distinct phases. In topological phase, there are two edge
states with imaginary energies $\pm i\gamma $, as verified in Fig. \ref{Fig4}%
. In the presence of a lattice vacancy, in both phases a state with purely
imaginary energy $+i\gamma $ ($-i\gamma $) localizes near the vacancy if $%
i_{0}\in B$ ($A$) sublattice, as shown in the energy distribution in the
insets (L1) and (R1) of \ref{Fig4}(b). In topological phase, when the
vacancy site $i_{0}\in B$, because of $t_{a}<t_{LR}$, the localized state
extends to $B$ site on the right, as confirmed by the density distribution
in the inset (L2) of \ref{Fig4}(b). While in trivial phase, due to $%
t_{a}>t_{LR}$, the localized state extends to $B$ site on the left, as shown
in the inset (R2) of \ref{Fig4}(b). If $i_{0}\in A$, the extension direction
of the localized state is opposite to that when $i_{0}\in B$. Due to the
particle-hole symmetry, the unpaired bound state with $E=\pm i\gamma $
cannot acquire a finite real energy through any perturbations with $\mathcal{%
C}$ symmetry, but may only change its imaginary part. This robust pining to
zero real energy is protected by $\mathcal{C}$ symmetry. Here, we also would
like to remark that the vacancy-induced localized states are robust to $%
\mathcal{PT}$ \ symmetry breaking, as it does to topological phase
transition (see Appendix \ref{A1} for more details).

\section{Photonic Graphene}

\label{S4}

\begin{figure}[tbp]
\centering\includegraphics[width=0.46\textwidth]{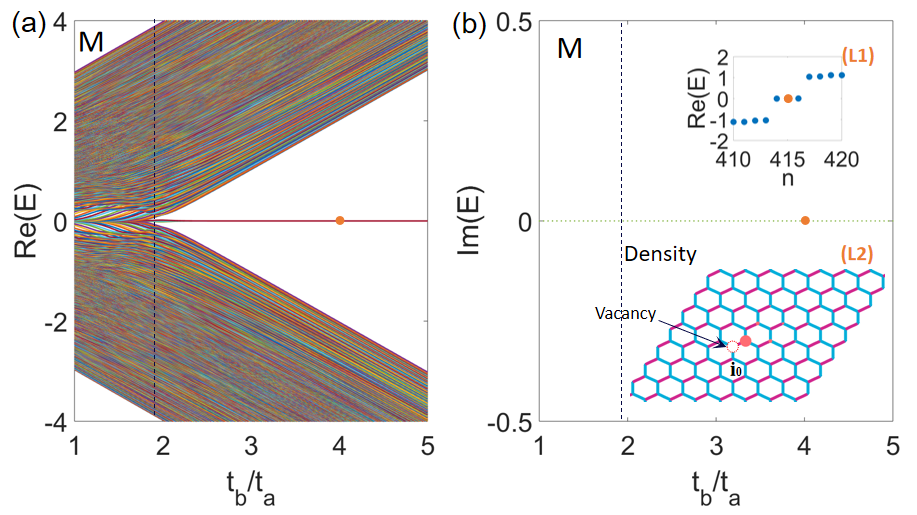}
\caption{Real (a) and complex (b) spectrums of graphene model with chiral
symmetry versus parameter $t_{b}/t_{a}$. The orange disk indicates
zero-energy state localized near the vacancy in the gapped phase. The top
inset $(\mathrm{L1})$ shows the real part of energies when $t_{b}/t_{a}$=4.0
and bottom one $(\mathrm{L2})$ gives density distribution of the localized
zero mode. The density is proportional to the radius of the pink spots.
Parameters are chosen as $\protect\lambda =0.2$, $t_{a}=1.0$, $\protect%
\gamma =0.$}
\label{Fig6}
\end{figure}
In this section, we consider the 2D honeycomb lattice sketched in Fig. \ref%
{Fig5}, which is relevant to photonic graphenes \cite{Oztas2018,Plotnik2014,
Efremidis2002,
Peleg2007,Sepkhanov2007,Treidel2008,Bartal2005,Polini2013,Treidel2010}. The
Bloch Hamiltonian on the honeycomb lattice reads $H_{G,0}\left( k\right)
=h_{0,x}\sigma _{x}+\left( h_{0,y}+i\lambda \right) \sigma _{y}+i\gamma
\sigma _{z},$where $h_{0,x}=t_{b}+2t_{a}\cos \left( 3k_{x}/2\right) \cos
\left( \sqrt{3}k_{y}/2\right) $, $h_{0,y}=-2t_{a}\sin \left( 3k_{x}/2\right)
\cos \left( \sqrt{3}k_{y}/2\right) $. Uneven hopping amplitudes introduce
the non-Hermitian term $i\lambda \sigma _{y}$, and the balanced gain/loss
gives rise to $i\gamma \sigma _{z}$. In the absence of non-Hermitian terms,
i.e., $\lambda =\gamma =0$, it corresponds to an isotropic graphene if $%
t_{a}/t_{b}=1$. As $\left\vert t_{a}/t_{b}\right\vert $ decreases, $C_{3}$
symmetry is broken, and the two Dirac nodes of vorticity $\pm \pi $
gradually approach, and finally meet up and annihilate at a time-reversal
invariant momenta at $\left\vert t_{a}/t_{b}\right\vert =1/2$. As $%
\left\vert t_{a}/t_{b}\right\vert $ decreases further, the system enter a
gapped topological phase, dubbed \textquotedblleft high-order topological
insulator\textquotedblright , which hosts zero-energy corner\ modes \cite%
{Motohiko2019, Motohiko2018}. In the presence of non-Hermitian term $%
i\lambda \sigma _{y}$ or $i\gamma \sigma _{z}$, the corner modes remain as
shown in Fig. \ref{Fig6} and \ref{Fig7} and this will be detailed in the
following.

\subsection{Chiral Symmetry Protected Modes}

In the absence of gain and loss, i.e., $\gamma =0$, this system has chiral
symmetry. The zero-energy corner modes localize at the corner. In addition,
the chiral symmetry ensures the eigenvalues appear in $\left(
E_{k},-E_{k}\right) $ pairs. Similar to analysis for SSH model in previous
section, we introduce a disordered local potential on one site. By gradually
varying the local potential as the same in Sec. \ref{S3} A, a bound state
also evolves from the bulk states and localizes near the defect. We
numerically compute the quantum states of this honeycomb lattice model ($%
32\times 26$ lattice) with vacancy site $i_{0}$ under open boundary
conditions. Figs. \ref{Fig6} (a) and (b) present the real and imaginary
parts of energies of the states, respectively. There are two distinct
phases, namely, metallic phase (M) and gapped phase. We find, in addition to
the two zero-energy corner modes, a localized zero mode (solid orange dot in
(L1) of Fig. \ref{Fig6}) appears near the vacancy, as shown in the inset
(L2).

\subsection{Particle-Hole Symmetry Protected Modes}

\begin{figure}[tbp]
\centering\includegraphics[width=0.46\textwidth]{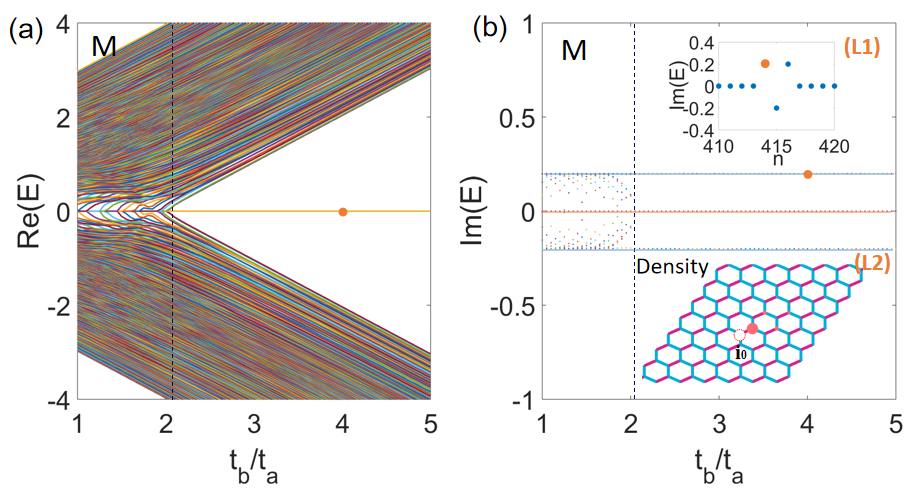}
\caption{Similar as Fig.~\protect\ref{Fig6} but with modified non-Hermitian
parameters $\protect\lambda =0$ and $\protect\gamma =0.2$.}
\label{Fig7}
\end{figure}

If $\lambda =0$, the particle-hole symmetry is respected. Because of the
eigenvalue correspondence ($E\leftrightarrow -E^{\ast }$), the unpaired
state must have exactly zero-energy or a purely imaginary energy. We repeat
the numerical processes and the results are plotted in Fig. \ref{Fig7}. In
the gapped phase, there are two corner states with imaginary energies $\pm
i\gamma $. The lattice vacancy induces an extra state (indicated by the
orange disk) with purely imaginary energy $+i\gamma $ ($-i\gamma $)
localized near the vacancy if $i_{0}\in B$ ($A$) sublattice. In the gapped
phase, when the vacancy site locates at $i_{0}\in A$, because $t_{a}<t_{b}$,
the localized state extended to $B$ site on the right, which is verified by
numerics in bottom inset (L2) of Fig. \ref{Fig7} (b). However, if $i_{0}\in A$,
the extension direction would be opposite, similar to the non-Hermitian SSH
model.

\section{Discussion and conclusion}

\label{S5}

In the presence of multi-vacancies, there exist a \textquotedblleft parity
effect", which states that for a system with odd number of vacancies, there
always exists a symmetry protected mode due to the eigenvalue
correspondence; while for system with even vacancies, the localized states
would possess a finite energy shift due to quantum tunneling effects. A numeric investigation on this matter is discussed in
Appendix \ref{A2}. In this paper, we mainly study one- and
two-dimensional systems. The general theory is also applicable to
three-dimensional lattice systems, such as the diamond lattice model. In
fact, the obtained result is applicable not only for the bipartite-lattice
models, but also for the lattice models with unit cell of even sites
preserving chiral or particle-hole symmetry \cite{Benalcazar2017}. These
conclusions can also be generalized to Hermitian systems with chiral or
particle-hole symmetry, where the zero mode gives rise to fractional charge
\cite{He2013}. The non-Hermitian SSH model and graphene model may be
realized by optical lattices, and the vacancy-induced localized modes could
be detected with current experimental techniques.

In summary, we have studied the vacancy-induced localized modes in
non-Hermitian systems with either chiral or particle-hole symmetries. The
localized states are symmetry-protected in the sense they are robust against
perturbations respecting the underlying symmetries.

\begin{acknowledgments}
This work is supported by NSFC under the grant No. 11504285, and the
Scientific Research Program Funded by Natural Science Basic Research Plan in
Shaanxi Province of China (Program Nos. 2018JQ1058 and 2019JM-001), the
Scientific Research Program Funded by Shaanxi Provincial Education
Department under the grant No. 17JK0805, and the scholarship from China
Scholarship Council (CSC) (Program No. 201708615072).
\end{acknowledgments}

\appendix

\section{Robustness to $\mathcal{PT}$ symmetry breaking}

\label{A1}
\begin{figure}[tbp]
\centering\includegraphics[width=0.46\textwidth]{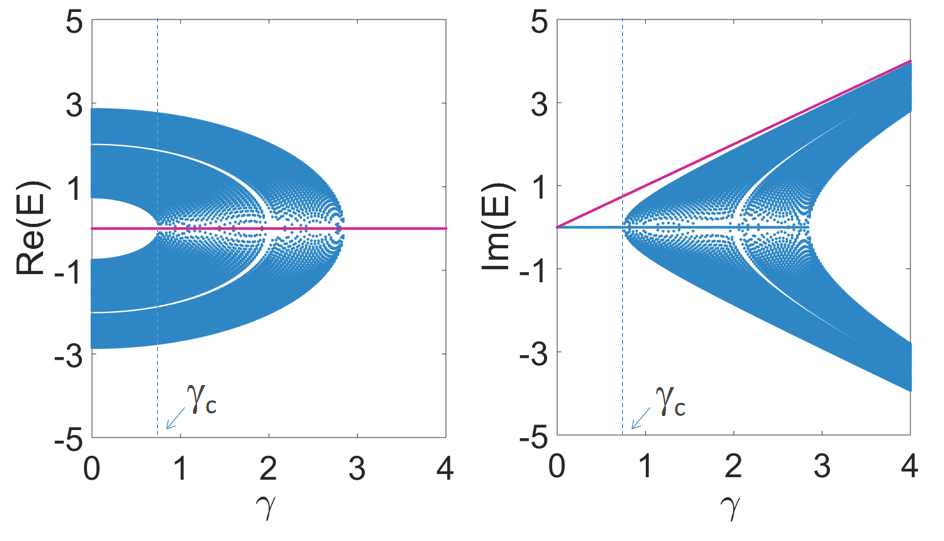}
\caption{The real part Re(E) and imaginary part Im(E) of eigenenergies of
SSH model with single lattice vacancy versus gain/loss strength $\protect\gamma $%
. Other parameters are fixed as $t_{a}=1.8$, $t_{LR}=1$, $t_{b}=0.1$ and $%
\protect\lambda =0$.}
\label{FigA1}
\end{figure}

Generally, the exceptional point is crucial for understanding many important
physical phenomena in non-Hermitian systems and it happens when the system
experiences a spontaneous symmetry breaking. In main text, we focus on the
symmetry-protected modes induced by local potentials at fixed on-site
gain/loss strength. To illustrate the role of $\mathcal{PT}$ symmetry
breaking and exceptional points, we study the spectrum through varying
gain/loss strength $\gamma$, as shown in Fig. \ref{FigA1}. It showcases the
system undergoes a $\mathcal{PT}$ symmetry breaking at the exceptional point
$\gamma _{c}$, where the bulk spectrum turns from real to imaginary as shown in Fig. \ref{FigA1}.
However, we find that any non-zero $\gamma $ would render a localized mode
with purely imaginary energy, as indicated by the red line in Fig. \ref%
{FigA1}, due to the particle-hole symmetry. So, the vacancy-induced
localized states are robust to $\mathcal{PT}$ \ symmetry breaking, as it
does to the topological phase transition.

\section{Parity effect in the presence of mutli-vacancies}

\label{A2}

\begin{figure}[tbp]
\centering\includegraphics[width=0.42\textwidth]{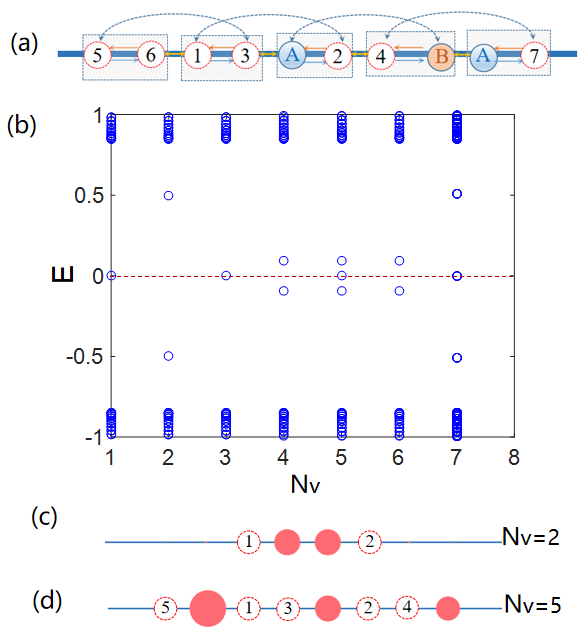}
\caption{(a) The SSH model with multi-vacancies. (b) The eigenenergies of SSH
model versus the the number of vacancies. $N_{\mathrm{v}}=m$ corresponds to
that there exist vacancies at sites $1,2,...,m$, as shown in (a). (c) and
(d) showcase the particle density distribution of localized modes in the
presence of $N_{\mathrm{v}}=2$ and $5$ vacancies. The density is
proportional to the radius of the pink spots. Parameters are chosen as $%
t_{a}=1.4$, $t_{LR}=0.5$, $t_{b}=0.1$, $\protect\lambda =0.1$, $\protect%
\gamma =0$, $N_{u}=100$.}
\label{FigA2}
\end{figure}

Without loss of generality, we take the chiral symmetric SSH as an example
to illustrate the parity effect regarding ``multi-vacancies''. In the presence of multi-vacancies, as
illustrated in Fig. \ref{FigA2} (a), the symmetry protected localized mode
exhibits parity effect.

Firstly, in the presence of odd vacancies ($N_{\mathrm{v}}=1,3,5,7$), there
always exists a localized zero-energy mode guaranteed by the chiral
symmetry, as confirmed by Fig.~\ref{FigA2} (b). Fig. \ref{FigA2} (d)
showcases the particle density distribution of localized modes with odd ($N_{\mathrm{v}}=5
$) vacancies in real space. Besides the zero-energy localized mode (indicated by the
bigger pink spot), there are also two localized in-gap modes with finite
energy (indicated by the two smaller pink spots, less localized than the
zero-energy mode). Secondly, for the system with even vacancies, the
tunneling effect could give rise to an energy splitting, so the zero-energy
state may disappear. For instance, the case of $N_{\mathrm{v}}=2$
demonstrates this point, as shown in Fig. \ref{FigA2} (b). However, the
in-gap modes possessing finite energy may also localize near vacancies, as
shown in Fig. \ref{FigA2} (c) in the case of even ($N_{\mathrm{v}}=2$)
vacancies.

In particle-hole symmetric non-Hermitian systems with multi-vacancies, the
parity effect also exists in analogy to that in aforementioned chiral
symmetric systems. The particle-hole-symmetry protected localized mode with
zero or purely imaginary energy always exists in the presence of odd
vacancies; while for system with even vacancies, the localized states would
possess a finite energy shift due to tunneling effects.

\newpage \clearpage\onecolumngrid\appendix

\end{document}